# Unlocking Social Media and User Generated Content as a Data Source for Knowledge Management

James Meneghello[1], Nik Thompson[2], Kevin Lee[3], Kok Wai Wong[4], Bilal Abu-Salih [5]

[1] Optika Solutions

[2] Curtin University

[3] Nottingham Trent University

[4] Murdoch University

[5] The University of Jordan

## ABSTRACT

The pervasiveness of Social Media and user-generated content has triggered an exponential increase in global data volumes. However, due to collection and extraction challenges, data in many feeds, embedded comments, reviews and testimonials are inaccessible as a generic data source. This paper incorporates Knowledge Management framework as a paradigm for knowledge management and data value extraction. This framework embodies solutions to unlock the potential of UGC as a rich, real-time data source for analytical applications. The contributions described in this paper are threefold. Firstly, a method for automatically navigating pagination systems to expose UGC for collection is presented. This is evaluated using browser emulation integrated with dynamic data collection. Secondly, a new method for collecting social data without any a priori knowledge of the sites is introduced. Finally, a new testbed is developed to reflect the current state of internet sites and shared publicly to encourage future research. The discussion benchmarks the new algorithm alongside existing data extraction techniques and provides evidence of the increased amount of UGC data made accessible by the new algorithm.

Keywords: Data Acquisition, Web Data Extraction, Data Manipulation, Content Discovery, Social Mining, User-Generated Content.

## 1. Introduction

User-generated content (UGC) is a promising data source for analytical applications, providing up-to-date information about a wide range of topics. There is increasing focus on the use of UGC on Social Networking Sites (SNS) to drive applications for emergency management (Buscaldi & Hernandez-Farias, 2015; Robinson, Power, & Cameron, 2013; Sakaki, Okazaki, & Matsuo, 2010) and opinion mining (Maynard, Bontcheva, & Rout, 2012), amongst other uses. While previous research has extensively analysed data from micro-blogging SNS such as Twitter, only a few studies have used traditional SNS (Wandhöfer et al., 2012) or embedded UGC such as user comments and reviews (Cao, Liao, Xu, & Bai, 2008). Since the shift to Web 2.0 paradigms, static content on websites has increasingly been replaced by UGC (Tim, 2005) that is not easily or generically accessible, resulting in a large amount of UGC being unavailable for mining.

While this data is timely and relevant, its relative inaccessibility is a major obstruction to its use in analytics. There are few methods of retrieving UGC embedded in pages, and there are no unifying interfaces to perform data collection from such sites. This contrasts to UGC on SNS, which is usually accessible via Application Programming Interfaces (APIs) - interfaces specifically made for automated applications to query stored data. Developing a method of accessing UGC in a standardised and reliable way would allow for the use of this data in social analytics applications, and provide the same accessibility as API-provided data from SNS.

Currently, several research efforts are undertaken to handle and manage the large scale of UGC in the quest for added value (Chan et al., 2018; Salih & Rahman, 2018). UGC presents an excellent opportunity for providing data about a wide range of topics, both historical and real-time. Users often post timely, relevant information on news articles about current events that can enhance or enable real-time event detection. Similarly, there is great value to be found in product-related UGC attached to product pages or reviews for marketing purposes(Tang, Ni, Xiong, & Zhu, 2015; Tuten & Solomon, 2017). With no formal interface to collect this data, generic collection of UGC is difficult and a significant proportion of data remains inaccessible, particularly if the data is on sites not accessible to web crawlers (Liakos, Ntoulas, Labrinidis, & Delis, 2016).

Web Data Extraction (WDE) algorithms have had success in extracting template-based data from web pages (Arasu & Garcia-Molina, 2003; Ferrara, De Meo, Fiumara, & Baumgartner, 2014; B. Liu, Grossman, & Zhai, 2003; Xia, 2009; Zhai & Liu, 2005). Similar techniques could be applied to extract UGC, but there are several complications related to current advancements in page design and rendering techniques. Logical page design and rendering on the internet has changed significantly since many WDE algorithms were developed, requiring newer structure detection techniques (Blanvillain, Kasioumis, & Banos, 2014). Modern pages are often dynamically rendered using Asynchronous Javascript and XML (AJAX), which many web scraping tools and WDE algorithms do not support. Complex user interactivity has also become the norm for modern sites, and needs to be accounted for.

UGC differs substantially to the data typically collected by WDE algorithms. Most of these algorithms are designed to extract discrete repeating structures - search record results (B. Liu et al., 2003), e-commerce data tables (Zhai & Liu, 2005) or the data from product detail pages (Crescenzi, Mecca, & Merialdo, 2001). UGC differs in comments often have replies that may or may not follow a similar tag structure, and can be nested to many depths. Thus, there may be nested data, but not always predictable nested structures. Any algorithm designed to extract UGC needs to consider and mitigate these problems.

This paper presents a complete data extraction framework able to automatically navigate pagination structures, expansion links and render dynamic DOM trees. The aim of this framework is to unlock the value of UGC stored within web pages and to address the inadequacy of incorporating WDE algorithm specifically for UGC. This framework is established to maximise the amount of useful data that may be accessed. UGC structures can then be isolated and filtered to provide a standardised interface for reliably accessing UGC on web pages. This study incorporates the Knowledge Management (KM) framework as a data analytics paradigm(Jennex & Bartczak, 2013). KM framework facilitates the implementation of the embodied filters, processes, and technologies constructed to satisfy the aim of the conducted data analytics and actionable intelligence.

This paper is structured as follows; Section 2 provides an overall background to the current state of the art approaches in UGC extraction. Section 3 describes the constructed framework and its embodied techniques.

Section 4 explains data generation and steps followed to collect UGC data followed by the mechanism carried out for information discovery and extraction as depicted in Section 5. Section 6 describes the experimental techniques conducted to verify the effectiveness and applicability of the constructed framework. A conclusion is provided in Section 7 that summarises the key contributions of this paper and indicating limitations and anticipated future research.

## 2. BACKGROUND

Since the emergence and proliferation of Web 2.0, the role of web browsers has changed to enable users to send and receive content by means of several online tools that commenced with e-mail applications chat, and chat forums that evolved into more recent and revolutionary electronic platforms such as OSNs(Abu-Salih, Wongthongtham, & Yan Chan, 2018; Wongthongtham & Salih, 2018). OSNs such as Facebook®, Twitter®, LiveBoon®, Orkut®, Pinterest®, Vine®, Tumblr®, Google Plus®, Instagram ® etc, which enable users to share videos, photos, files and instant conversations. These platforms provide an important means by which communities can grow and consolidate, allowing individuals or groups to share concepts and visions with others(Abu-Salih, Wongthongtham, Chan, & Zhu, 2018). Moreover, in addition to playing an active and distinctive role as effective media of social interaction, these OSNs allow users to become acquainted with and understand the cultures of different peoples.

WDE algorithms are designed to extract semi-structured data from web pages have been in development for the last 15years (Arasu & Garcia-Molina, 2003; Atzeni & Mecca, 1997; Cao et al., 2008; Crescenzi et al., 2001; B. Liu et al., 2003; Miao, Tatemura, Hsiung, Sawires, & Moser, 2009; Zhai & Liu, 2005). WDE algorithms have had great success in scraping search result pages in order to facilitate web crawling (B. Liu et al., 2003), as well as collecting e-commerce data (such as competitor prices) (Zhai & Liu, 2005) database detail pages (Thamviset & Wongthanavasu, 2014a), and news articles (Gupta, Mittal, Bishnoi, Maheshwari, & Patel, 2016).

WDE algorithms use wrappers (programs that extract data from a source into a relational form) to extract data from pages, but the method of creating wrappers has evolved over time. Initially, wrapper development was manual and required a developer to write appropriate code to extract the desired fields from a single page (Atzeni & Mecca, 1997; Crescenzi & Mecca, 1998) Most wrappers are not generic, and each new data source required the development of a new wrapper. As this represented a significant development and maintenance cost, automated techniques were highly sought after.

Automated WDE algorithms can be classified based on their method of function and the manner of data input. Functionally, these algorithms work either by comparing the page DOM trees (Ferrara & Baumgartner, 2011; Jindal & Liu, 2010; B. Liu et al., 2003; Sleiman & Corchuelo, 2014; Zhai & Liu, 2005) by examining and mining the text directly (Thamviset & Wongthanavasu, 2014a, 2014b), or by recognising visually-similar page regions (W. Liu, Meng, & Meng, 2010; Wei Liu, 2005). Some algorithms also combine several of these techniques to improve accuracy, such as using text mining to locate data and tree comparisons to identify the enclosing environment (Thamviset & Wongthanavasu, 2014a). Similar techniques can be used to scrape UGC from pages, thus UGC extraction can therefore be considered a subfield of Web Data Extraction - henceforth, User Generated Content Extraction (UGCE).

Prior research in which UGC has been extracted from web pages have used bespoke wrappers to parse page contents (Cao et al., 2008; Mishne & Glance, 2006). Perhaps due to time constraints, most prior studies using UGC were limited to platforms that already provide an API for data access. For example, Twitter feeds have been used to detect earthquakes (Sakaki et al., 2010) and predict depression in social media users (De Choudhury, Gamon, Counts, & Horvitz, 2013). The data input for these projects is of course constrained by the presence and functionality of a vendor supplied API, in this instance, Twitter. Some studies have accessed web pages as part of social data analytics, but had no automated method of extracting the data from pages (Meneghello, Thompson, & Lee, 2014).

UGC is plentiful on the internet, but only API-provided content is easily accessible, comprising only a fraction of total UGC. By developing an automated method to extract UGC in a standardised format, web pages such as news articles, blogs and other widely used media can be unlocked and made accessible as viable data source. This would greatly broaden the scope of data available for use in social analytics applications and decision making.

## 3. User-Generated Content Extraction

UGC can represent current events, peoples' views and thoughts on issues or their current environment (Subrahmanyam, Reich, Waechter, & Espinoza, 2008) . These comments, statuses, messages and tweets, although given many different names, often possess a common set of attributes. While different sources may provide more or less metadata for a social event, the minimal set usually includes an origin, a timestamp and some message content. An origin can be a person, an account or commonly a pseudonym, while a timestamp can be a date, a time, a relative time or any combination of these. Content usually contains the detail about the event occurring - a message between friends, a comment on a topic under discussion or an update on the state of an origin. Figure 1 shows an example of UGC, with each user comment containing each of these attributes - name, date and content.

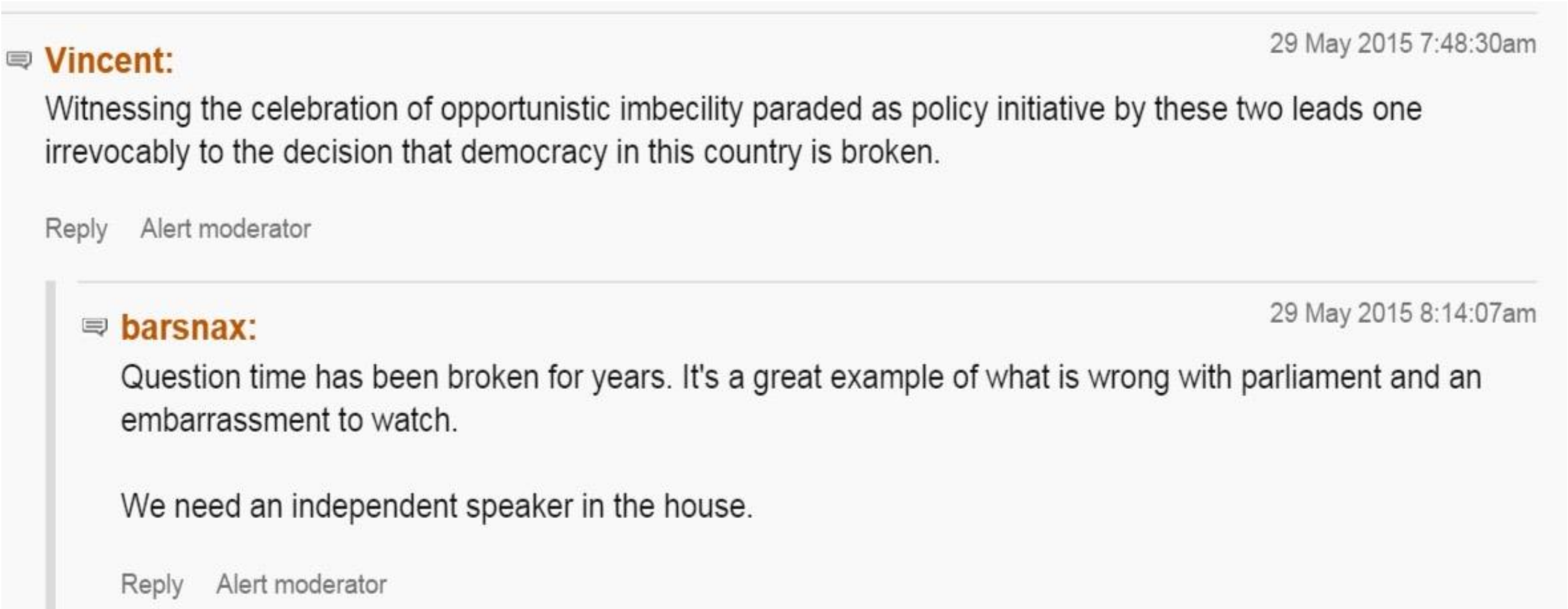

*Figure 1. Embedded UGC from a popular news website, ABC Australia.*

The UGC present on such pages is similar to that on SNS, except it tends to be anchored around a particular subject or content. This makes page-based UGC extremely useful for user based analytics such as intent or sentiment analysis. UGC is already major consideration for users during e-Commerce research (Cheong & Morrison, 2008), often having significant effect during product selection. While data from SNS is frequently used for this, inaccessibility of page-based UGC is an obstacle to providing more effective social analytics. The proposed approach aims to address this deficiency by developing a reliable, accessible method of automatically extracting UGC from pages.

**UGC Extraction - System Architecture**

The wealth of UGC presents a unique opportunity for organisations to obtain the excessive use of such data abundance to increase their revenues. Hence, there is an imperative need to capture, load, store, process, analyse, transform, interpret, and visualise such manifold social datasets to develop meaningful insights that are specific to an application's domain. This can be achieved by incorporating KM framework leveraging the proliferation of UGC to help companies to integrate their internal processes, the UGC's data lake and KM system to infer valuable insights and to provide actionable intelligence. Hence, this research incorporates Knowledge management framework (Jennex, 2017; Jennex & Bartczak, 2013) to provide a systematic approach for UGC extraction and analysis. Figure 2 illustrates the system architecture that includes the main phases followed in this research incorporating KM framework.

As depicted in Figure 2, three primary steps have been identified that must be considered when developing an automated UGC extraction (UGCE) process:

- **Data Collection and Acquisition:** a method of automatically interacting with and collecting data from dynamic sources without any *a priori* knowledge of the source structure;
- **Information Discovery and Extraction:** an algorithm to detect commonly-occurring data structures and extract location and data types of fields contained within, while avoiding nested data complications, and A rule-based filter that ensures only UGC is extracted from the page, which enhances precision.
- **Actionable Intelligence:** the aforementioned steps establish the solid ground to conduct proper data analytics leveraging the UGC extraction approach.

This paper tackles the first two phases (viz**.** Data Collection and Acquisition, and Information Discovery and Extraction). Actionable intelligence phase will be addressed in our future research. The following sections detail the developed UGC extraction approach.

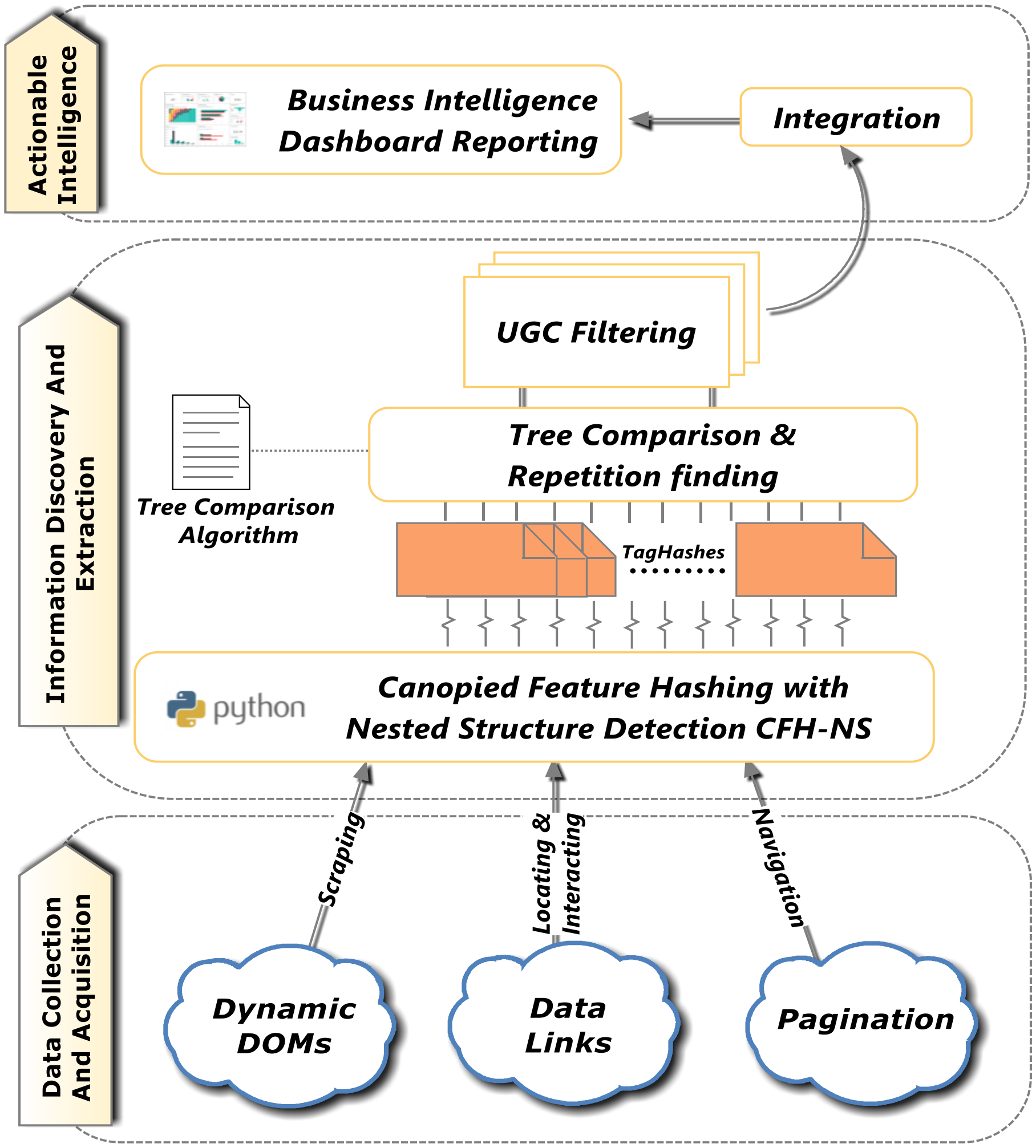

*Figure 2: System Architecture incorporating KM Framework*

## 4. DATA COLLECTION AND ACQUISITION

Prior to running any data extraction algorithms on a data source, significant preparation is required. Because AJAX and dynamic DOM manipulation are in heavy usage on modern sites, it is no longer possible to simply issue an HTTP request and save any raw page data returned. The advent of the "single page website" requires that any method of extracting UGC needs to fully emulate a user by emulating the browser rendering, JavaScript engine and any required user interaction in order to expose linked UGC hidden behind pagination fields.

**Accessing Deep Data**

In order to interact with these elements, they first need to be reliably identified. Figure 3 depicts the preparation process. Once the page and all AJAX requests have been completed and rendered, expansion and pagination elements can be identified. This process must be performed recursively as new content injected dynamically into the DOM from an AJAX request may itself contain further expansion or pagination elements, which must also be processed. A copy of each rendered page is cached. Once this process has been completed, saved page content may be passed to any extraction algorithm. Thus the data collection step is compatible even with legacy data extraction algorithms not designed to operate on dynamic DOM trees.

Pagination serves to limit the size of requests required to dynamically render pages, but also to restrict the amount of content available to the user. While a user may not want to view 600 comments at once, applications analysing social comments desire access to every available piece of data, and therefore requires a method of identifying and triggering pagination and expansion elements so that comprehensive data can be collected.

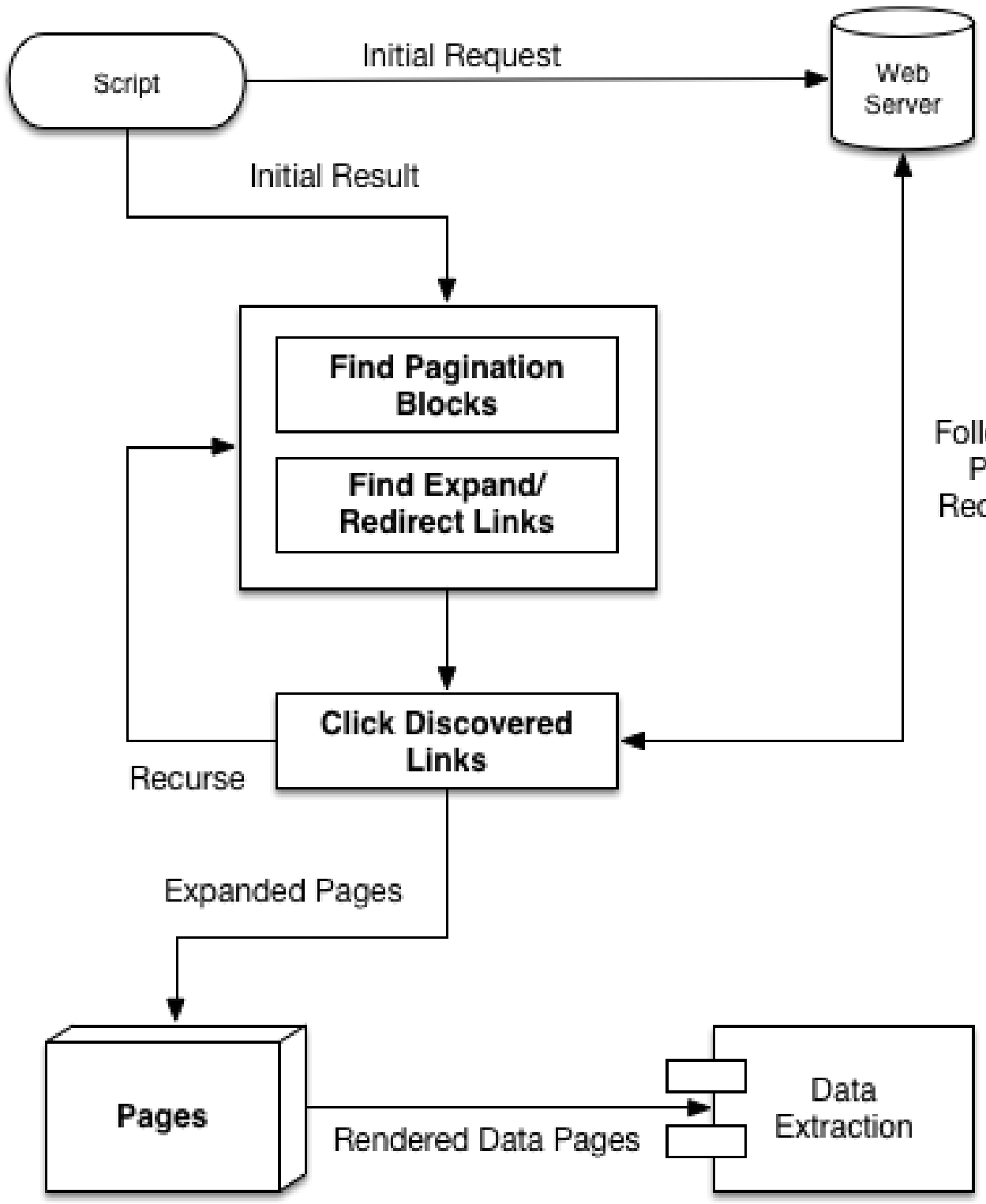

*Figure 3: Proposed architecture for pagination handling*

Fortunately, the frequent use of Cascading StyleSheet (CSS) and Javascript frameworks has led to similarities between pagination structures and expansion links. By designing broad rules, most of these structures can be identified regardless of page design. Three primary structures have been identified that require handling:

- **Pagination:** Pagination elements generally consist of a numeric sequence containing links to data pages, sometimes within a select box, as seen in Figure 4.
- **Expansion Link:** An expansion link is often interleaved throughout nested comment trees, and can be clicked by a user to expand the replies within that thread.
- **Redirection Link:** A redirect link is used on some pages when comments are completely isolated from the page, and provides a link to another page consisting primarily of social content.

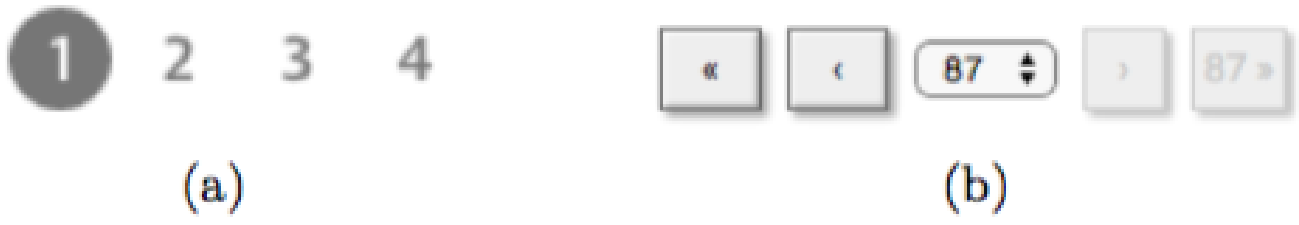

*Figure 4: Examples of pagination elements taken from popular news websites and online forums*

To effectively collect the raw data required to extract UGC, the collection process must be robust and able to properly handle the above structures. Each part of this process is described in the following sections.

## 4.1 Scraping Dynamic DOMs

To retrieve the fully-rendered page, a method of emulating user interaction and standard browser processes (such as Javascript and page rendering) is required. Solutions have previously been suggested that use an embedded browser to render page content before further processing takes place (Mesbah, Van Deursen, & Lenselink, 2012; Xia, 2009). Automated approaches have no need for actual presentation of the page, so a screen-less solution is preferable in order to reduce external dependencies and resource usage.

To satisfy these requirements, PhantomJS was employed(Hidayat, 2013). PhantomJS is a headless browser emulator capable of dynamically rendering pages, including handling any requisite Javascript requests.

## 4.2 Locating and Interacting with Data Links

In order to extract all possible data from a page, user interaction with pagination and expansion elements needs to be simulated. The interaction should cause the browser emulator to dynamically modify the DOM tree, which can then be stored for later parsing by extraction algorithms.

To facilitate this interaction emulation process, Selenium (Holmes & Kellogg, 2006), a real-time interactive web testing framework, was used. Selenium is able to emulate user interaction while rendering pages. Combined, PhantomJS and Selenium provide a virtual browser and testing agent that is able to load and render content and interact with it in a realistic manner, while still providing common DOM manipulation and parsing tools such as XPath and CSS selectors.

**Discovering Data Links**

Content expansion elements can be identified by searching for elements containing certain text patterns, which allows a reasonable estimation of redirection and expansion links to be made. Links that lead to additional data are often identified by a qualifier followed by a certain noun, such as “more comments”. While there are several English words synonymous to “more”" and “comments'”, a regular expression is able to match combinations of these words in order. Similar principles can also be applied to languages other than English.

```
COMMENT_MULTIPLE_WORDS = ['read', 'more', 'all', 'show']
COMMENT_LINK_WORDS = ['comments', 'replies']

COMMENT_WORD_REGEX = regex.compile(
  r'.*{}{}.*|.*{}.*'.format(
    r'(?={})'.format(
      r'|'.join(
        [r'.*\b{}\b'.format(word) for word in
         ↪  COMMENT_MULTIPLE_WORDS]
      )
    ),
    r'(?={})'.format(
      r'|'.join(
        [r'.*\b{}\b'.format(word) for word in
         ↪  COMMENT_LINK_WORDS]
      )
    ),
    r'(\d+\b({}))'.format(r'|'.join(COMMENT_LINK_WORDS))
  ), regex.I
)

""" Generates:

.*
(?=.*\bread\b.*\bmore\b|.*\ball\b|.*\bshow\b)
(?=.*\bcomments\b|.*\breplies\b)
.*
|
.*(\d+\b(comments|replies)).*

"""
```

*Figure 5: Locating and Interacting with Data Links*

The example Python code presented in Figure 5 will generate a regular expression that can be applied to a page to locate link candidates. The candidates can then be filtered further to exclude any words that can lead us to unwanted content. Of the discovered elements, those that are able to respond to user interaction are stored.

Once appropriate link elements have been discovered, Selenium is used to simulate a user “click"” on these elements. The resulting data from this interaction, either a new page or dynamically-loaded content, is cached in the set of page content for later use.

## 4.3 Navigating through Pagination

Identifying and interacting with pagination elements on a page is more difficult than the data links in the previous section. While some pagination blocks contain names or classes that can be matched to regular expressions, many structures do not contain relevant identifiers. Thus, to detect these blocks, the algorithm instead searches for features that may suggest the presence of a pagination structure.

Figure 6 presents a basic algorithm designed to locate these structures. Pagination blocks typically consist of a list of elements containing consecutive ordered integers representing data pages. Because values are often skipped for presentation, the entire list is unlikely to be consecutive. Shortened lists like this may be included by instead searching for spans, e.g. 1,2,3,7,15 becomes 1-3,7,15. These numbers can either be the values of individual clickable elements or option fields within a select box, which are both common designs for pagination blocks. Structures that contain value lists following these rules are located, as these identify pagination blocks. These blocks can also contain buttons labelled “previous”, “next” or “all”, which can be accessed by the algorithm in the same way that a real user would. Repeatedly interacting with the “next” button cycles through every available page, while the “all” button often redirects to a view showing all available data.

```
function FindPaginationNumbers(domtree)
    for each tag in domtree do
        values ← tag.children.strings
        digits ← LIST(value for value in values if value is an inte-
ger)
        spans ← INTSPAN(digits)      ▷ converts [1,2,3,5] into [1-3,5]
        if length(digits) > length(spans) then              ▷ found
consecutive ints
            if SORTED(digits) == digits then      ▷ ints appeared in
order
                links ← buttons or links in tag.children
                if length(links) ⩾ length(digits) then
                    pagination ← tag
                    return pagination
                end if
            end if
        end if
    end for
end function
```

*Figure 6: simplified example of finding a numbered pagination block*

The pagination block can be expressed as an identifying XPath (Meneghello, 2015) which allows the block to be located again easily on new pages. Links such as “next” or “all” are expressed as direct relative XPath, while numbered elements are expressed as grouped relative XPath. These XPath are then passed to Selenium, which handles the interaction.

Using this process, it is possible to automate interaction with sites for the purpose of gathering UGC. Redirection links are identified and copies of the dynamically-rendered page content are saved. Next, expansion links are recursively interacted with to expand the DOM tree with additional UGC. During this process, Pagination structures are identified and expanded to include additional UGC pages. Upon completion of this process, a full set of page content is available. This is suitable to undergo the Data Extraction process presented in the next section.

## 5. INFORMATION DISCOVERY AND EXTRACTION

CFH-NS (Canopied Feature Hashing with Nested Structure Detection) is a new UGC extraction algorithm developed to specifically target user-generated content. The example implementation of this algorithm utilizes the Python (Van Rossum & Drake, 2003) programming language and BeautifulSoup (Richardson, 2013), a DOM tree parsing library.

To parse page structures, the algorithm operates by breaking a DOM tree into individual branches, anchoring at different nodes within the tree. An example DOM tree for a single social comment is presented in Figure 7 in which the desired anchor would be the top-level LI tag. To locate similar structures, the algorithm attempts to match the structure derived from that branch with other branches in the tree. A quick and effective means of comparing these branches is through the generation of a canopied feature hash, or TagHash.

```
<li>
    <a id="m_ucMessageDisplay1548290_m_anchMessageAnchor"
     ↪  name="m1548290"></a>
    <h3 class="">Patrick:</h3>
    <p class="date">29 Jan 2015 3:15:55pm</p>
    <p>Abbott displays all the hallmarks of a highly
     ↪  delusional right-man. He appears egotistical in the
     ↪  extreme and it should now be obvious to all that he
     ↪  is an extremely dangerous individual and one who
     ↪  should never be in a position of power, let alone
     ↪  being leader of a nation</p>
    <p>
        <span>
            <a class="popup"
             ↪  href="NewMessage.aspx?b=69&amp;t=12532">
                Reply
            </a>
        </span>
        <span>
            <a class="popup"
             ↪  href="AlertModerator.aspx?b=69&amp;m=1548290">
                Alert moderator
            </a>
        </span>
    </p>
    <ul></ul>
</li>
```

*Figure 7: An Example DOM tree for a single comment*

### 5.1 Canopied Feature Hashes

TagHashes express the structure of a tree and its associated tags in a simplified manner, stripping out unique features. Rather than attempting to match element attribute values directly (which may include unique attributes, such as an ID) the *existence* of attributes is matched, specifically ID and name. This provides a level of differentiation not present if only matching tag types, and assists in tree comparison.

TagHashes are expressed internally as a Python dictionary and are serialised as a Javascript Object Notation (JSON (Crockford, 2006)) string, which are virtually identical in practice. JSON is used due to simple bidirectional serialisation, and Python dictionaries have the additional benefit of supporting tree comparison functions. Figs. 8a and 8b present two example TagHashes of the same structure, Figure 7 to different depths. Each TagHash is a simplified representation of the DOM tree for that structure, genericised to allow for partial-tree matching whilst retaining some differentiating features. Each TagHash is effectively a signature for the DOM block.

```
{'li.00': [
  {'a.11': []},
  {'h3.00': []},
  {'p.00': []},
  {'p.00': []},
  {'p.00': []},
  {'ul.00': []},
]}
```

(a) Depth 1

```
{'li.00': [
  {'a.11': []},
  {'h3.00': []},
  {'p.00': []},
  {'p.00': []},
  {'p.00': [
    {'span.00': []},
    {'span.00': []},
  ]},
  {'ul.00': []},
]}
```

(b) Depth 2

*Figure 8: TagHashes of the Structure in Fig. 7*

To match similar structures together and identify as many UGC structures as possible, a method of comparing DOM structures is required. Several techniques have been used in previous research (Bille, 2005; Jindal & Liu, 2010; Zhai & Liu, 2005). Initially, a simple tree-comparison algorithm was created that checked for an identical match between structures. Due to the structurally-variable nature of UGC and page design (such as the use of BR or P tags to structure paragraphs), this was later refined to allow for partial matches.

TagHashes are compared using text comparison tools. By using a text representation of each TagHash and computing the difference between them using standard tools surprisingly good results can be achieved, including the ability to handle unaligned branches in the trees being compared. The string representations of Python dictionaries (and JSON) ensure that the type of difference (whether an attribute change, a full branch change or a node change) is easily distinguishable. This allows for granular control of similarity scores for different trees, similar to how tree edit-distance comparisons (Bille, 2005) function.

Figure 9 shows the result of a text diff between two TagHashes. As illustrated, a node addition or subtraction is represented by a line starting with "+" or "-" respectively, with accompanying braces. A tag modification is represented by two lines starting with "+" and "-" and no accompanying braces. The addition or subtraction of a branch is represented similarly to a node addition or subtraction, but contains multiple tags on the diff line. Checking the number of tags on a diff line can determine how many tags made up the branch and score appropriately. The diff format provides an efficient way of performing scored comparisons by assigning each action (tag modification, tag addition/subtraction, branch addition/subtraction) with a penalty score. By comparing this penalty with the number of tags involved in the comparison, it is possible to assign a score or similarity percentage between two trees - if over a threshold, the two trees are deemed similar. An appropriate threshold can be designated depending on application requirements for signal-to-noise: a higher threshold will match only very similar structures and can potentially miss data from some sites with less structured designs.

```
{'li.00': [
  {'div.01': []},
  {'div.00': []},
  {'p.00': []},
  {'a.00': []},
  {'div.00': []},
  {'a.00': []}
]}
```

(a) TagHash 1

```
{'li.00': [
  {'div.00': []},
  {'p.00': []},
  {'dl.00': []},
  {'p.00': []}
]}
```

(b) TagHash 2

```
--- a
+++ b
{
  'li.00': [
  @@ -0,3 +0,5 @@
  +{'div.01': []},
   {'div.00': []},
   {'p.00': []},
    {
   +'a.00': [],
   -'dl.00': [],
   },
    {
   +'div.00': [],
   -'p.00': [],
   },
  +{'a.00': []}
  ],
}
```

(c) Diff

*Figure 9: Example diff: two TagHashes and their resulting diff, showing node and branch modifications*

## 5.2 Finding Repeating Structures

To detect repeating UGC structures on a page, a list of TagHashes is first built - a TagHashList. This can be built using every possible tag on a page, or restricted to a subset of tag types likely to hold UGC items. The TagHashList class contains additional functionality that automatically groups together structures into buckets based on their TagHash, and can then output common structural XPaths (Meneghello, 2015). The TagHashList also provides the ability to apply custom filters over the data to ensure that the identified structures contain certain items - this is particularly useful when dealing with UGC.

With a populated TagHashList, structural XPath can be generated to match discovered structures. These XPath can then be constructed into wrappers and used to locate structures to extract data from additional pages.

## 5.3 UGC Filtering

Upon conclusion of the structure-matching process, a set of similar data structures remains, but little information as to what they contain. As UGC is the primary focus of this process, it is desirable to be able to filter the detected data records to only those that are likely to contain UGC. These same principles can be applied to other data types as well – for example filters might be developed to limit the data records to just those containing e-commerce data.

As discussed previously, UGC can be viewed as a structure that contains at least three primary elements: an origin, a timestamp and a message. In terms of datatypes, this is a string (origin), a date/time/datetime (timestamp) and text (message). To expand the applicability of filters, fields within the discovered structures can be typed and filters designed to match these types.

### 5.3.1 Type Discovery

Determining the datatype of a single field is a non-trivial task. A set of three words could be a short string, or a short block of text. For UGC, the difference is important as a string can represent a user identifier or location, while text usually indicates message content. To ensure that fields are not mis-classified, the individual fields must not be examined in isolation, but rather grouped with the matching fields from other structures to provide context.

With multiple values available per-field, the *probability* that a field is of a certain datatype can be determined and compared to a predetermined threshold. Using this method of probability-based type-checking, the reliability of field data typing can be improved.

To perform this type determination, value of each descendant tag of the parent structure is first extracted. By iterating through the tree below the parent, all descendant tags can then be extracted. This presents a further challenge: if a UGC structure is nested within a parent structure, such as replies to a comment, fields can begin to be accessed from within the nested structure. To ensure that this does not occur, the TagHash of possible nested children is compared against their ancestors - if a match exists, a nested structure has been discovered. This approach does not work for nested children that have a different structure to parents, however.

Some types require more work to validate than others. Text can be discovered by checking for the presence of newline characters or BR tags and having high average word count, while strings are short and generally do not contain punctuation. Dates can be particularly problematic, as they can be represented in standard ways (e.g. ISO8601 (Klyne & Newman, 2002)) or just as times (e.g. 5:12am) or even in relative terms (e.g. 16 minutes ago). Very lenient date parsing must be used, and the probability-based checking can be used to exclude some fields that contain dates embedded within text content - while some content fields may contain a time or date, it's not likely that every instance will.

The type determinations for each instance of the field are represented as boolean values in a list, and a simple mean calculation is used to determine the percentage of instances that identify as each type. If any of these types exceeds the predetermined threshold, the field is considered to be of that type. Once an appropriate type has been decided, this may be flagged based on the relative field location and the position of the data (e.g. within a certain attribute, or the text value of the tag).

During this process, additional data alignment and cleaning is also performed. If the field is only aligned with a few structures, it may be optional, and gets flagged as such. Fields that have constant templated values are discarded. Any tag that contains a series of text-like tags is noted as a text parent, and children are discarded from the list - this prevents individual paragraphs being included as separate fields.

### 5.3.2 Field Cleansing

Figure 10 presents a completed set of type-checked fields. With each valid field sorted into type-buckets, a filter checking for the existence of certain types can be applied. For UGC, a string field, a datetime field and a text field are required, all of which are present in the Figure 10. For this approach, the structure being tested satisfies the requirements of a UGC field and is appended to a list of discovered fields.

```
{
  'datetime': ['p[1]/inner_text'],
  'url': [
    'p[3]/span[2]/a[1]/@href',
    'p[3]/span[1]/a[1]/@href'
  ],
  'text': ['p[2]/text'],
  'string': ['h3[1]/string']
}
```

*Figure 10: Field types discovered from a set of UGC structures similar to Figure 7.*

Using the discovered set of UGC structures and their fields, data is extracted and aligned. In this phase of filtering, the most appropriate instance of each type is automatically selected. Once completed, the resulting structure and list of fields is generated into a wrapper, seen in Figure 11.

```
{
  '// li [ count(a)=1 and count(h3)=1 and count(p)=3 ]': {
    'other': [],
    'content': ['p[2]/text'],
    'datetime': ['p[1]/inner_text'],
    'name': ['h3[1]/string']
  }
}
```

*Figure 11: A wrapper constructed from an expanded version of Figure 7, including fields.*

### 5.3.3 Filtered Extraction

This wrapper can then be used to extract data from other pages made from the same templates. Figure 12 below shows the data ultimately extracted using the wrapper generated in the previous step. The output data contains the UGC that has been detected, which has now been datatyped and structured consistently. The content is injected into a structure with appropriately named fields. This data is now in a form that may be easily imported and used in a variety of applications or stored for later use after any required processing or cleaning.

```
[
  {
  'other': [],
  'content': ['Abbott displays all the hallmarks of a highly
    ↪ delusional...'],
  'datetime': ['29 Jan 2015 3:15:55pm'],
  'name': ['Patrick']
  },
  {
  'other': [],
  'content': ["Every footy team needs a head-kicker but you
    ↪ don't make him captain"],
  'datetime': ['29 Jan 2015 3:47:38pm'],
  'name': ['Tony']
  },
  {
  'other': [],
  'content': ['@Tony:\nTony Abbott displayed all of his
    ↪ head kicking prowess as...'],
  'datetime': ['29 Jan 2015 4:17:03pm'],
  'name': ['JohnC']
  },
  {
  'other': [],
  'content': ['Like'],
  'datetime': ['29 Jan 2015 6:07:58pm'],
  'name': ['Arthur']
  },
]
```

*Figure 12: Data extracted from a document using the wrapper generated in Figure 11.*

## 6. Evaluation

The effectiveness of the new CFH-NS algorithm was evaluated by comparing it with similar algorithms from previous research. To test these algorithms, each was run against a testbed of HTML data and extracted structures were counted to determine recall and precision. To evaluate the effectiveness of automated interaction algorithms, the amount of data collected using traditional HTTP requests was compared with the data collected using the interaction and rendering techniques.

Testbeds for web data extraction algorithms already exist, such as TBDW (Yamada, Craswell, Nakatoh, & Hirokawa, 2004), ViNTs (Wei Liu, 2005) and others. Each of these testbeds contains a set of pages that contain data records, such as Search Record Results or List Records. Evaluating an algorithm against these testbeds is relatively simple - the algorithm is run against the pages and the number of detected data records is compared with the number that is known to exist on each page.

Two testbeds commonly-used in previous research (TBDW and ViNTs) were compiled in 2004 and 2005, and represent typical web design techniques for that period. However, web design has dramatically changed in the time since then. HTML standards and design paradigms have evolved to encourage better structure while introducing higher complexity, which significantly alters the requirements for web data extraction algorithms. Modern pages also rely heavily on dynamic DOM interaction to render content, making the collection mechanisms implemented in many WDE algorithms non-functional.

These changes have effectively rendered the standard evaluation testbeds inappropriate, as they no longer represent the state of design on the internet. Regardless, they would be unsuitable for evaluation in this context: user-generated content was seldom embedded into pages during the time period in which the testbeds were compiled. The only sites that tended to contain UGC were message boards or forums, whereas the shift to Web 2.0 ideologies has driven the expansion of primarily user-generated content pages. These fundamental changes required not only the development of new extraction techniques, but also of new evaluation techniques including a test bed. This test bed and supporting data has contributed to the public domain knowledge to support future research and evaluation and to enable researchers to directly compare their findings with the results described in this manuscript.

## 6.1 Methodology

To test the CFH-NS algorithm against existing solutions, a new testbed was compiled that is more representative of the current state-of-the-art in web design and development methodologies.

A set of candidate URLs was crowdsourced, with participants directed to provide sites that they commonly visited that used UGC in some way. The candidate URLs provided covered news sites, social media, blogs and online stores. This crowd sourced list was then filtered to exclude sites that met any of the below criteria:

a) The presence of an inappropriate amount of UGC (i.e less than 10 items or more than 1000 items).
b) Multiple sites with duplicate structures, e.g. blogs operating on the same software with similar design themes.
c) Obscure sites or those that are unrepresentative of the state of the internet, e.g. sites that haven't undergone design upgrades in the last decade.
d) Sites on unreliable servers that didn't always return page data in a timely fashion.

After filtering, a total of 49 URLs remained. The pages require various amounts of interactivity to retrieve data, consisting of different types of pagination and expand/redirect link types and covering a wide range of design strategies for these elements. The page designs also vary, with some using tag types in structures (such as list elements), and others preferring to use DIVs for everything (which can make extraction significantly more difficult). The site designs represent modern web design, and are representative of current and popular content on the internet. These include a mixture of sites that require dynamic interaction or statically presented content.

While the older testbeds from prior research are unsuitable for use in this evaluation, the algorithms themselves are not. Although designed to detect simple data records such as search results and e-commerce data tables, UGC constructs share many of the same qualities and legacy WDE algorithms should be able to detect (but not necessarily isolate) UGC. For these, only the effectiveness of structure detection is evaluated.

Furthermore, some of the algorithms from prior research come with their own data collection processes that are not able to handle dynamic content. Where this was the case, rendered dynamic data was directly supplied to those algorithms, to allow for a fairer side-by-side comparison.

## 6.2 User Generated Content Extraction

Web Data Extraction algorithms come in two broad types: those designed to operate on single pages, and those designed to operate on sets of similar pages. As the CFH-NS algorithm is designed to operate on a single page, it was benchmarked against other algorithms of this type.

For this evaluation, an attempt was made to acquire the code for several WDE algorithms by directly contacting the authors of previous research, however limited replies were received. Few algorithms are provided online, and a number of those are now non-functional. Hence, the algorithms being evaluated are Data Extraction based on Partial Tree Alignment (DEPTA) (Zhai & Liu, 2005) and Bottom-Up Wrapper (BUW) (Thamviset & Wongthanavasu, 2014a), both of which are publicly available.

CFH-NS is able to split up fields into their appropriate type and identify fields relevant to UGC, which neither DEPTA nor BUW was designed to do, so this functionality was not evaluated. The DEPTA and BUW algorithms are evaluated slightly differently. Neither algorithm provides the ability to filter UGC, so this process was performed manually upon the various record types returned by each algorithm. If a discovered record represented a single social comment that contained the necessary, it was deemed a successful extract. Any records that contained nested data, did not contain the necessary fields or represented a field within a record were considered a failed extract. Records that were unrelated to UGC were discarded and not included in the evaluation, and are not reflected in recall or precision.

DEPTA provides a Java binary that can be operated on a set of collected data, but BUW only provides a live web interface and performs its own collection. DEPTA was passed a full set of post-rendered data and interacted content, but BUW was unable to perform these duties on its own. To account for these problems, two sets of tests were completed. Using data procured, interacted and rendered by CFH-NS, the performance of both CFH-NS and DEPTA was evaluated for extracting UGC structures. Then, using their own collection mechanisms, DEPTA and BUW were again evaluated. In keeping with evaluation practices in previous research, results for recall and precision are provided: “All Results” describes recall and precision if including collection or WDE failures as part of the results, while “Success Only” excludes failures and only determines recall and precision for pages in which at least one result was found.
Recall represents the number of relevant records available on a page that were successfully discovered, while precision represents the number of discovered elements that were relevant. F-score is calculated using a standard balanced algorithm that represents the harmonic mean of precision and recall:

$$F_1 = 2 \cdot \frac{\text{precision} \cdot \text{recall}}{\text{precision} + \text{recall}}$$

These results are presented in Figure 13. Actual Results-Paginated (AR-P) is the number of available results from the paginated dataset, produced by the new CFH-NS algorithm. Actual Results (AR) is the number of available results without any interaction required. The figures in each algorithm's column represent the total number of results found and the number of relevant results. DEPTA-L and BUW-L are the “live” versions of each algorithm, and represent the results if the algorithms perform their own data collection rather than using CFH-NS's collected set.

| | AR-P | CFH-NS | DEPTA | AR | DEPTA-L | BUW-L |
|---|---|---|---|---|---|---|
| Total | 3155 | 3454/2981 | 853/764 | 1675 | 149/124 | 704/593 |
| **All Results:** | | | | | | |
| Recall | | **94.5%** | 24.2% | | 7.4% | 35.4% |
| Precision | | 89.2% | **89.6%** | | 83.2% | 84.2% |
| F-score | | **0.92** | 0.38 | | 0.14 | 0.50 |
| **Success Only:** | | | | | | |
| Recall | | **94.5%** | 48.1% | | 48.1% | 71.6% |
| Precision | | 89.2% | **89.6%** | | 83.2% | 84.2% |
| F-score | | **0.92** | 0.63 | | 0.61 | 0.77 |

*Figure 13: Content extraction experimental results*

As seen in the table, CFH-NS provides a significant increase in recall compared to DEPTA using the paginated set, while retaining similar precision. This also represents a significant improvement in both recall and precision compared to BUW, though BUW performs admirably when only counting successful tests. When comparing all results, CFH-NS represents a major improvement in F-score.

A large part of the difference in recall between CFH-NS and the live versions of DEPTA and BUW is due to dynamic DOM parsing. Neither algorithm is able to render the DOM dynamically, and are both thus unable to retrieve the UGC on pages. By counting the URLs in which both DEPTA and BUW fail to retrieve data, it is possible to determine how many pages use dynamic DOM injection to render social content. 51% of sites tested require dynamic DOM parsing to retrieve UGC, indicating that algorithms without the ability to deal with this problem are unable to collect significant amounts of UGC.

CFH-NS represents a major improvement in the ability to collect UGC from web pages. This technique can be used to develop a reliable API for accessing previously-inaccessible UGC, broadening the scope of social data analytics applications. This is particularly useful for advertising and product-based analytics, as CFH-NS can be used to generically mine product reviews and user comments in the context of the product. It can also be used to mine real-time blogs and microblogs, providing a broader set of data sources for event detection.

## 6.3 Dynamic Interaction

To test the effectiveness of automatic dynamic interaction, the number of records available on a page without interaction is compared against the number of records available using interaction. Some URLs in the dataset normally used pagination, but there was not enough UGC present to be paged. Of 49 URLs, 10 (20.5%) contained data hidden behind pagination or expansion links. In addition, 19 (38.8%) had pagination code, though there was not enough UGC on the selected pages to trigger its usage. In total, 29 (59.2%) sites required navigation of pagination to extract available data. This proportion of sites requiring complex user interaction to access data highlights the need to develop better methods of automated interaction, a field that has received little attention to date.

| URL | AR-P | AR | Increase |
|---|---|---|---|
| http://newsfeed.gawk... | 42 | 20 | 210.0% |
| http://the-toast.net... | 41 | 27 | 151.9% |
| http://www.adelaiden... | 101 | 50 | 202.0% |
| http://www.cracked.c... | 592 | 34 | 1741.2% |
| http://www.dailytele... | 72 | 50 | 144.0% |
| http://www.destructo... | 219 | 50 | 438.0% |
| http://www.dpreview.... | 193 | 176 | 109.7% |
| http://www.smh.com.a... | 47 | 10 | 470.0% |
| http://www.theage.co... | 545 | 10 | 5450.0% |
| http://www.tripadvis... | 20 | 10 | 200.0% |
| http://www.theguardi... | 86 | 20 | 430.0% |
| Totals | 1958 | 457 | 428.4% |

*Figure 14: Additional data retrieved through CFH-NS's automated interaction techniques*

These pages and the results of automated interaction are presented in Figure 14. As shown in the table, automated interaction yielded 428% more UGC from these pages than without using interaction, a substantial increase. This suggests that a significant volume of web data is currently inaccessible to many existing data collection mechanisms.

Interestingly, sites that require automated interaction are very often those that use pagination to handle due to heavy traffic or an overabundance of UGC. These may indeed be the most useful and desirable candidates for UGC mining. Being able to access this previously-unavailable data using CFH-NS is a significant step forward, and allows for more extensive mining of UGC to unlock its analytical value.

## 6.4 Limitations

The three techniques presented in this paper have some limitations that would provide a good basis for future work.

The automated interaction algorithm is subject to certain phrases or design styles being used in the source code of the pages, and does not function well outside these general cases - a more robust method of detecting expansion and redirection links could provide access to significantly more data.

The CFH-NS algorithm does not perform well with very loose tag structures, particularly those that rely heavily on inline HTML tags (such as B, P and SPAN tags). It was developed to work with well-designed page structures, which causes performance to suffer on legacy designs. Combination of the algorithm with those that perform text-based context extraction (such as BUW (Thamviset & Wongthanavasu, 2014a)) could improve recall and precision on certain sites.

Finally, the probabilistic datatype determination technique created is a simple threshold-based model and could benefit from the use of more complex statistical methods, though the accuracy of the current model performs adequately for this implementation.

## 7. Conclusion

This paper incorporates KM framework as a paradigm for knowledge management and data value extraction. The incorporated framework presents three contributions to the field of UGC and social media analytics. The first contribution involves a method for automatically navigating generic sites to extract UGC. This overcomes the limitations of traditional web data extraction methods by fully emulating the actions of a user, thus revealing data elements hidden by Javascript or pagination. The second contribution, the CFH-NS algorithm makes it possible to discover and extract nested UGC page structures, while classifying UGC fields. Thirdly, as the research revealed the need for a more modern testbed, this has been developed and shared in the public domain to invite and support future research. All contributions described in this paper have been empirically evaluated, both with existing testbeds if applicable, as well as with the new testbed described above.

CFH-NS, the UGC extraction algorithm, represents a new algorithm capable of automatically extracting UGC from web pages. Results demonstrate that overall data recall is significantly improved compared to existing data extraction algorithms. The user emulation and interaction techniques can significantly increase the amount of UGC collected from web pages by automatically navigating UGC pagination while the data typing and classification algorithms provide a standardised interface to the UGC.

The combined use of these techniques increases the reach of UGC mining into pages containing less structured content, including news articles and comments, forums and other sites that do not provide an API for retrieving data. This data is made accessible through a standard interface that enables more direct integration into analytical applications.

There are many opportunities for future work on these techniques. This includes; (i) implementing the actionable intelligence layer indicated in the KM framework;(ii) improving recall ability and more precise selection of data; and (iii) using intelligent sourcing algorithms to locate new UGC sources. This work could then be combined into a full framework designed to integrate diverse social data sources, and provide a unified interface for performing social analytics. This is the ultimate objective: to realize the full potential of UGC and social data as a low-cost and real-time data source, capable of providing actionable insights on a global scale.